\documentclass[aps,prd,preprint,showpacs,nofootinbib]{revtex4}
\usepackage{latexsym,bm,amsmath,amssymb,xfrac,xcolor}
\DeclareMathAlphabet{\mathpzc}{OT1}{pzc}{m}{it}

\newcommand{\KN}{\mathbin{\bigcirc\mspace{-15mu}\wedge\mspace{3mu}}}
\begin{document}

\title{Holographic c-theorem and Born-Infeld Gravity Theories}
\author{G{\"o}khan Alka\c{c}}
\email{alkac@metu.edu.tr}
\author{Bayram Tekin}
\email{btekin@metu.edu.tr}
\affiliation{Department of Physics, Faculty of Arts and  Sciences,\\
             Middle East Technical University, 06800, Ankara, Turkey}

\date{\today}

\begin{abstract}
The requirement of the existence of a holographic c-function for higher derivative theories is a very restrictive one and hence most theories do not possess this property. Here, we show that, when some of the parameters are fixed, the $D\geq3$ Born-Infeld gravity theories admit a holographic c-function. We work out the details of the $D=3$ theory with no free parameters, which is a non-minimal Born-Infeld type extension of new massive gravity.   Moreover, we show that these theories generate an infinite number of higher derivative models admitting a c-function in a suitable expansion and therefore they can be studied at any truncated order.
\end{abstract}
 \maketitle
\section{Introduction}
Einstein's theory of gravity needs to be modified at UV (and IR) but it is very hard to do this modification while keeping its nice properties intact. Salient features of the theory, for example, the fact that it has a unique vacuum [(flat ($\Lambda_0=0$), de Sitter ($\Lambda_0>0$) or anti-de Sitter ($\Lambda_0<0$)] and that it has a single massless unitary spin-2 particle about its vacuum, no longer hold for generic modifications such as adding higher order terms in the contractions of the Riemann tensor. For example, even a simple modification of the form $R-2\Lambda_0+\alpha R^2+\beta R_{\mu\nu}^2$ induces a massive spin-2 ghost, a massive spin-0 particle and degeneracy of the vacuum for generic values of $\alpha$ and $\beta$. Given this state of affairs, it was a rather surprising result to see that there is a family of Born-Infeld (BI) type theories in $D>3$ dimensions that have the same features as Einstein's gravity but with improved high energy behavior \cite{Gullu:2014gza,Gullu:2015cha}. Explicit construction of these theories requires a rather complicated analysis since in constant curvature backgrounds one needs to consider the contributions of infinitely many higher order terms in principle \cite{Gullu:2010em}. Therefore, in what follows, we shall only give a brief description of these theories.  The defining action is given by
\begin{equation}
S_{\text{BI}}=\frac{2m^2}{\gamma\,\ell_P^{(D-2)}}\int d^{D}x\sqrt{-g}\,\left[\sqrt{\det\left(\delta^{\mu}_\nu+\frac{\gamma}{m^2} A^{\mu}_\nu\right)}-\left(1+\frac{\gamma\Lambda_{0}}{m^2}\right)\right],\label{Generic_BI}
\end{equation}
where $\gamma$ is a dimensionless parameter (the BI parameter)\footnote{We introduce the dimensionful parameter $m^2$, which makes the parameter $\gamma$ defined in \cite{Gullu:2014gza} dimensionless. } and $\ell_P$ is the Planck length. The \emph{minimal} version of the A-tensor in generic $D$-dimensions must be at least quadratic in the Riemann tensor and its contractions. There are seven possible linearly independent terms and hence seven {\it a priori} free parameters (not counting $\gamma$). However, in generic $D$ dimensions, three of these parameters do not survive when the uniqueness of the vacuum, the unitarity of the massless spin-2 mode about this vacuum and the condition that there are no other perturbative modes are imposed. This eventually reduces the tensor $A_{\mu\nu}$ to
\begin{align}
A_{\mu\nu}= & R_{\mu\nu}+\beta S_{\mu\nu}+\frac{\gamma}{m^2}\left(a_{1}\mathcal{W}_{\mu\nu}+a_{2}C_{\mu\rho\nu\sigma}R^{\rho\sigma}+\frac{\beta+1}{4}R_{\mu\rho}R_{\nu}^{\rho}+a_{4}S_{\mu\rho}S_{\nu}^{\rho}\right)\label{eq2} \\
 +&\frac{\gamma g_{\mu\nu}}{m^2D}\left[\left(\frac{\left(D-1\right)^{2}}{4\left(D-2\right)\left(D-3\right)}-a_{1}\right)\mathcal{W}-\frac{\beta}{4}R_{\rho\sigma}^{2}+\left(\frac{\beta\left(\beta+2\right)}{2}+\frac{D\left(4-3D\right)}{4\left(D-2\right)^{2}}-a_{4}\right)S_{\rho\sigma}^{2}\right]\label{eq:Main_result_for_Amn_0}\nonumber
\end{align}
with  $\mathcal{W}_{\mu\nu}\equiv C_{\mu\rho\alpha\beta}C_{\nu}^{\phantom{\nu}\rho\alpha\beta}$ and $\mathcal{W}\equiv g^{\mu\nu}\mathcal{W}_{\mu\nu}$.
$C_{\mu\alpha\nu\beta}$ is the Weyl tensor and $S_{\mu\nu}=R_{\mu\nu}-\frac{1}{D}g_{\mu\nu}R$ is the traceless Ricci tensor. There are several important points to note here. For the choice ${\gamma\Lambda_{0}}{m^2}=-1$,  the second term in (\ref{Generic_BI}) vanishes and the field redefinition $g_{\mu\nu}\rightarrow g_{\mu\nu}-\frac{\gamma}{m^2} A_{\mu\nu} $ yields a trivial Lagrangian. Therefore, this particular point is excluded from the parameter space of the theory and so we assume ${\gamma\Lambda_{0}}{m^2}\neq-1$. The dimensionless parameters $\beta$, $a_1$, $a_2$ and $a_4$ are four free parameters compatible with the above-mentioned requirements for generic $D$-dimensions with two exceptions. In $D=4$, due to the identity $\mathcal{W}_{\mu\nu}=\frac{1}{4}g_{\mu\nu}\mathcal{W}$, which is valid for any smooth metric, the parameter $a_1$ drops out and one is left with three free parameters.  In $D=3$, the Weyl tensor vanishes and one just sets $D=3$ after excluding terms with the Weyl tensor and arrives at a theory with two free parameters  $\beta$ and $a_4$.

An expansion of the action (\ref{Generic_BI})  in $\sfrac{1}{m^2}$  yields Einstein $+$ the Gauss-Bonnet combination $+$ highly nontrivial terms which preserve all the nice properties of Einstein's theory even when the expansion is truncated at a particular order. The vacuum of this theory, which is given in (\ref{BIvac}), is unique when the unitarity of the massless spin-2 excitation is imposed, and at low energies, for the effective cosmological constant $\Lambda=-\frac{(D-1)(D-2)}{2L^2}\neq0$, the free part of the BI action is equivalent to that of the cosmological Einstein's theory with the effective Planck constant $\ell_P^\text{eff}$ given as
\begin{equation}
	\frac{1}{(\ell_P^\text{eff})^{D-2}}=\frac{1}{\ell_P^{D-2}}\left(1-\frac{\gamma\Lambda}{m^2}\right)\left(1+\frac{\gamma\Lambda}{(D-2)m^2}\right)^{D-2}.
\end{equation}

It is clear that the $D\rightarrow3$ limit of (\ref{eq2}) makes sense after dropping out the terms with the Weyl tensor. This yields the following A-tensor
\begin{align}
A_{\mu\nu}= & R_{\mu\nu}+\beta S_{\mu\nu}+\frac{\gamma}{m^2}\left(\frac{\beta+1}{4}R_{\mu\rho}R_{\nu}^{\rho}+a_{4}S_{\mu\rho}S_{\nu}^{\rho}\right)\nonumber \\
+&\frac{\gamma g_{\mu\nu}}{3m^2}\left[-\frac{\beta}{4}R_{\rho\sigma}^{2}+\left(\frac{\beta\left(\beta+2\right)}{2}-\frac{15}{4}-a_{4}\right)S_{\rho\sigma}^{2}\right].
\end{align}
For $D=3$, as we show explicitly in Section \ref{sec:D3BI},  the theory becomes a non-minimal BI-type extension of New Massive Gravity \cite{Bergshoeff:2009hq}.

Let us recall that in $D=3$ dimensions, there is a second BI-type gravity which started all the discussion in this context. It is given by the action \cite{Gullu:2010pc}
\begin{equation}
S_\text{BINMG}=-\frac{4m^{2}}{\ell_P}\int d^{3}x\sqrt{- g}\,\left[\sqrt{\det\left(\delta^{\mu}_{\nu}+\frac{\sigma}{m^{2}}G_\nu^\mu\right)}-\left(1-\frac{\Lambda_0}{2m^2}\right)\right],\label{3DBI}
\end{equation}
where $G_{\mu\nu}$ is the Einstein tensor and $\sigma=\pm 1$. The field redefinition $g_{\mu\nu}\rightarrow$ $g_{\mu\nu}-\frac{\sigma}{m^2}G_{\mu\nu}$ makes the theory trivial for $\Lambda_{0}=2m^2$. Therefore, this point is excluded from the parameter space. This theory has a massive spin-2 graviton with a mass square $m_g^2= -\sigma m^2 + \Lambda $ around its unique vacuum with the effective cosmological $\Lambda=\sigma \Lambda_{0} \left(1-\frac{\Lambda_0}{4m^2}\right)$ subjected to the constraint $\Lambda_0<2m^2$. It is rather interesting that, when truncated at ${\mathcal{O}}(\sfrac{1}{m^2})$, this theory reproduces Einstein's gravity and at ${\mathcal{O}}(\sfrac{1}{m^4})$ it reproduces the New Massive Gravity (NMG) \cite{Bergshoeff:2009hq} with its two vacua. Therefore, it is named as the Born-Infeld extension of NMG (BINMG). Furthermore, up to ${\mathcal{O}}(\sfrac{1}{m^8})$ it gives the theories given in \cite{Sinha:2010ai} which were obtained by demanding the existence of a holographic c-function and the theories given in \cite{Paulos:2010ke} at any order. Some holographic properties of the theory were studied in \cite{Gullu:2010st,Nam:2010dd} and it was also shown to appear as the counterterm in AdS$_4$ in \cite{Jatkar:2011ue}. Here, motivated by a possible holographic c-theorem in arbitrary dimensions  proposed in \cite{Myers:2010xs,Myers:2010tj}, which extends the construction of  \cite{Freedman:1999gp} to both odd and even dimensions, we initiate the study of the holographic properties of the massless BI theories for $D>3$. 

The layout of this paper is as follows: In section \ref{sec:BINMG}, we reconstruct the c-function of BINMG theory with a method that is useful for the study of generic BI-type gravities. Section \ref{sec:BI} is devoted to the BI gravities for $D\geq3$ where we construct a monotonically increasing function which might describe a holographic RG-flow by fixing two of the free parameters of the theory. In Section \ref{sec:D3BI}, we study the $D=3$ case, where there remains no free parameters, and show that the monotonic function takes the value of the central charge of the boundary field theory at the UV fixed point, by checking the coefficient of the Weyl anomaly.

\section{c-function in BINMG}\label{sec:BINMG}
To set the stage and notation, we start with Einstein's theory and introduce the relevant tools in constructing the c-functions. Then, we move on to NMG and BINMG in the subsequent parts.
\subsection{Einstein's Theory}\label{sec:Einstein}
Let us first see how the c-function is constructed for 3D Einstein's gravity defined by the action
\begin{equation}
S=\int d^{3}x\,\sqrt{-g}\left[\frac{1}{\ell_P}\left(R-2\Lambda_{0}\right)+\mathcal{L}_M\right].
\end{equation}
The matter Lagrangian $\mathcal{L}_M$ is chosen such that it will induce an RG-flow. We begin with the domain wall ansatz,
\begin{equation}
ds^2=e^{2A(r)}(-dt^2+dx^2)+dr^2,\label{dwall3d}
\end{equation}
and then impose the null-energy condition (NEC) for the matter fields, i.e, $T_{\mu\nu}\zeta^\mu\zeta^\nu\geq0$ for an arbitrary null vector $\zeta^\mu$. For the choice $\zeta^\mu=(\zeta^t, \zeta^r, \zeta^x)=v(e^{-A(r)},1,0)$, where $v$ is an arbitrary function,  the NEC becomes
\begin{equation}
v^2(e^{2A(r)}\,T_{tt}+T_{rr})\geq0, \qquad \text{or} \qquad T^t_t-T^r_r\leq0,
\end{equation}
where the latter form is more suitable for our purposes. Using the field equations,
 \begin{equation}
 G_{\mu\nu}+\Lambda_{0}\,g_{\mu\nu}=\ell_P\, T_{\mu\nu},
 \end{equation}
it is easy to see that the null-energy condition implies
\begin{equation}
T^t_t-T^r_r=\frac{1}{\ell_P}(G^t_t-G^r_r)=\frac{1}{\ell_P}A^{\prime\prime}\leq0,\label{NECeinstein}
\end{equation}
where $A^\prime(r)=\frac{dA}{dr}$.
This suggests that one can define the function
\begin{equation}
c(r):= \frac{1}{\ell_P A^\prime(r)},
\end{equation}
which is monotonically increasing since
\begin{equation}
	c^\prime(r)=-\frac{1}{\ell_P} \frac{A^{\prime\prime}}{{A^{\prime}}^2}\geq0.
\end{equation}
When there is no matter, the metric (\ref{dwall3d}) solves the field equations for $\Lambda_{0}=-\frac{1}{L^2}$ with  $A(r)=\frac{r}{L}$, which is the AdS spacetime with the AdS length $L$, and at the UV boundary ($r\rightarrow\infty$), the value of the function $c(r)$ is the central charge of the dual CFT\footnote{With the usual normalization of Brown and Henneaux \cite{Brown:1986nw}, one has $c=\frac{3L}{2G_3}$.} 
\begin{equation}
	c=\frac{L}{\ell_P}.
\end{equation}
Assuming that the spacetime asymptotes to AdS both in the UV ($r\rightarrow\infty$) and the IR ($r\rightarrow- \infty$) we have $c_{UV}\geq c_{IR}$, establishing the holographic c-theorem and fulfilling the expectation that the number of degrees of freedom in the UV regime of a field theory is larger than that of the IR regime.
\subsection{A Method to Derive NEC and NMG as an example}
In order to derive the NEC, we consider the minimal coupling of a free scalar field to gravity as\footnote{This method was recently used to constrain the Ricci polynomial and the Riemann cubic gravities in \cite{Li:2017txk}. For the study of the c-theorem in Horndeski gravity, see \cite{Li:2018kqp}.}  
\begin{equation}
S = \int d^{3}x\,\sqrt{-g}\left(\mathcal{L}^{gr} - \frac12 \partial_\mu\phi \partial^\mu\phi\right)\,\label{actionmatter},
\end{equation}
where $\mathcal{L}^{gr}$ is the Lagrangian of a generic gravitational theory. Assuming $\phi=\phi(r)$, with $r$ being the radial coordinate, the NEC becomes the statement of the positivity of the radial kinetic energy
\begin{equation}
	\ell_P(-T^{t}_t + T^{r}_{r}) = \frac12 \phi'^2\geq0.\label{NECmatter}
\end{equation}
We can find the left-hand side of this inequality directly from the action if we use the following domain wall  ansatz
\begin{equation}
ds^2=e^{2A(r)}(-dt^2+dx^2)+e^{2B(r)}dr^2,\label{dwall3dwithB}
\end{equation}
which yields our original metric (\ref{dwall3d}) for the gauge choice $B(r)=0$. Inserting this ansatz into the action (\ref{actionmatter}) gives an effective action for the functions $(A(r),B(r),\phi(r))$, from which the field equations can be found. After finding the Euler-Lagrange equations from the effective action, we fix the gauge by $B(r)=0$ and end up with three differential equations with two unknowns $(A(r),\phi(r))$. Note that the difference $-T^{t}_t + T^{r}_{r}$ in the NEC (\ref{NECmatter})  leads to a cancellation of the cosmological constant if we write the left-hand side of the inequality in terms of the gravitational field equations. Therefore, elimination of the cosmological constant in the equations obtained from the effective action gives the NEC in the form
\begin{equation}
  \frac{1}{2\ell_P}\phi'^2= F(A^\prime,A^{\prime\prime},\ldots)\geq0,\label{eq17}
\end{equation}
where $F$ is a smooth function of its arguments. For an arbitrary gravity theory, one can try to use this inequality to find a c-function as we did in the case of Einstein's theory in part \ref{sec:Einstein}.

As an example of this method, let us study the most general quadratic theory in 3D defined by
\begin{equation}
	S=\frac{1}{\ell_P}\int d^{3}x\,\sqrt{-g}\left[\sigma R-2\Lambda_{0}+\frac{1}{m^2}(\lambda_1 R_{\mu\nu}^2+\lambda_2 R^2)\right],\label{eq18}
\end{equation}
where $\lambda_1$ and $\lambda_2$ are arbitrary constants at this stage. This construction will lead to the result that NMG is the unique theory which possesses a c-function at this order, which was first found in \cite{Sinha:2010ai}.
The explicit form of (\ref{eq17}) for the action (\ref{eq18}) can be computed as
\begin{eqnarray}
 \frac{1}{2\ell_P} \phi'^2&=&-\left(\sigma+\frac{4}{m^2}(\lambda_1+3\lambda_2){A^{\prime}}^2\right)A^{\prime\prime}+\frac{4}{m^2}(3\lambda_1+8\lambda_2){A^{\prime\prime}}^2\nonumber \\
 &&+\frac{4}{m^2}(3\lambda_1+8\lambda_2)(2A^\prime A^{\prime\prime\prime}+A^{\prime\prime\prime\prime})\geq0.\label{NECquad}
\end{eqnarray}
In the limit $m^2\rightarrow \infty$, one recovers the condition for Einstein's gravity (\ref{NECeinstein}) as required. Our aim is to find a monotonically increasing function, and this can be achieved by eliminating the higher derivative terms with the choice $3\lambda_1+8\lambda_2=0$, which leads to NMG \cite{Sinha:2010ai}. With the choice $\lambda_1=1$ and $\lambda_2=-\frac{3}{8}$, the NEC (\ref{NECquad}) reduces to
\begin{equation}
	 \frac{1}{2\ell_P} \phi'^2=-\left(\sigma-\frac{1}{2m^2}{A^{\prime}}^2\right)A^{\prime\prime}\geq0.
\end{equation}
By inspection, we define the function \cite{Sinha:2010ai}
\begin{equation}
	c(r):=\frac{1}{\ell_P{A^\prime}} \left(\sigma+\frac{1}{2m^2}{A^{\prime}}^2\right),\label{21}
\end{equation}
which is monotonically increasing since
\begin{equation}
		c^\prime(r)=-\frac{1}{\ell_P{A^{\prime}}^2} \left(\sigma-\frac{1}{2m^2}{A^{\prime}}^2\right)A^{\prime\prime}\geq0.
\end{equation}
The AdS spacetime with $A(r)=\frac{r}{L}$ is again a solution of the field equations with no matter fields ($\phi(r)=0$) if the vacuum equation is satisfied as
\begin{equation}
	\Lambda_0 + \frac{\sigma}{L^2}-\frac{1}{4m^2L^4}=0,\label{NMGvacuum}
\end{equation}
which generically has two solutions for the effective cosmological constant $\Lambda=-\frac{1}{L^2}$. The central charge of the boundary theory follows from (\ref{21}) as 
\begin{equation}
		c=\frac{L}{\ell_P} \left(\sigma+\frac{1}{2m^2L^2}\right),\label{NMGc}
\end{equation}
which is first computed in \cite{Bergshoeff:2009aq}. One can continue this procedure to higher orders as was done in \cite{Sinha:2010ai} up to ${\mathcal{O}}(\sfrac{1}{m^8})$ and in \cite{Paulos:2010ke} up to arbitrarily high orders. For all these theories, the vacuum equation at order $n$, which  is a polynomial of $\Lambda$ of degree $n$, has generically more than one real root, destroying the uniqueness of the vacuum. The non-uniqueness of vacuum in gravity is a serious obstruction to the validity of the theory as one cannot decide which vacuum is viable using any physical arguments such as energy considerations as all these spaces have different asymptotics. This motivates the discussion of BINMG which has a unique vacuum.
\subsection{BINMG}
Let us consider the following 3D action:
\begin{equation}
S=-\frac{4m^{2}}{\ell_P}\int d^{3}x\sqrt{-\det g}\,\left[\sqrt{\det\left(\delta^\mu_\nu+\frac{\sigma}{m^{2}}A^\mu_\nu\right)}-\left(1-\frac{\Lambda_0}{2m^2}\right)\right],\label{3DBIgeneral}
\end{equation}
where
\begin{equation}
	A_{\mu\nu}=R_{\mu\nu}+\lambda\, g_{\mu\nu} R.
\end{equation}
and $\sigma=\pm 1$ is introduced to control the sign. For $\lambda=-\frac{1}{2}$ or  $\lambda=-\frac{1}{6}$, when expanded in powers of $\sfrac{1}{m^2}$, this action gives NMG at the quadratic order \cite{Gullu:2010pc}. However, at the third and fourth orders,  only for $\lambda=-\frac{1}{2}$, it gives rise to the theories which were found in \cite{Sinha:2010ai} by requiring the existence of a holographic c-theorem. For this choice, the full theory was also shown to admit a holographic c-function in \cite{Gullu:2010st} and this is the theory that we will refer to as BINMG. By a very simple observation, one can see that these properties are deeply connected.

Let us consider a Lagrangian given as an expansion of the form
\begin{equation}
	\mathcal{L}=\sum_{n=0}^{\infty} \left(\frac{1}{m^2}\right)^n \mathcal{L}^{(n)},
\end{equation}
which will lead to field equations as
\begin{equation}
	\Phi_{\mu\nu}=\sum_{n=0}^{\infty} \left(\frac{1}{m^2}\right)^n \Phi_{\mu\nu}^{(n)},
\end{equation}
where $\Phi_{\mu\nu}^{(n)}$ are the field equations corresponding the order-$n$ Lagrangian $\mathcal{L}^{(n)}$.
With this form of the field equations, the NEC becomes
\begin{equation}
		-T^{t}_t + T^{r}_{r} = -\Phi^{t}_t + \Phi^{r}_{r}  = \sum_{n=0}^{\infty} \left(\frac{1}{m^2}\right)^n \left(-\Phi^{t}_t + \Phi^{r}_{r}\right)^{(n)} = \frac{1}{2\ell_P} \phi'^2\geq0.
\end{equation}
We have seen that for the existence of a c-function, there should not be any terms with more than a second derivative. Therefore, if the full theory has a c-function, then there should not be any higher derivative terms at each order since there can be no cancellation between different orders. This naturally implies that there exists a theory admitting a c-function which is defined by the $n$th order Lagrangian $\mathcal{L}^{(n)}$. One can also take any linear combination of the Lagrangians $\mathcal{L}^{(n)}$'s to define theories with a c-function. As a result, since the BINMG theory has a c-function, it is natural for theories obtained in the $(\sfrac{1}{m^2})$-expansion to have a c-function.

Indeed, one can use this argument to study the BI-type theories as follows: one fixes the parameters in the theory by working out the first few Lagrangians in the $(\sfrac{1}{m^2})$-expansion. After fixing the parameters, it is much easier to check if the full theory admits a c-function or not. Since the full Lagrangian and the corresponding field equations are rather complicated, this is a very powerful method especially in the study of higher dimensional theories.

Let us see how this method works for our 3D example. We insert the ansatz (\ref{dwall3dwithB}) into the full Lagrangian (\ref{3DBIgeneral}) and study the leading terms. The NEC gives the following inequalities at the first two orders in $\sfrac{1}{m^2}$:
\begin{eqnarray}
	0&\leq& -\sigma (1+3\lambda)A^{\prime\prime}, \\ 
	0&\leq& -2(1+3\lambda)^2{A^{\prime}}^2A^{\prime\prime}+2(2\lambda+1)(6\lambda+1)A^{\prime}A^{\prime\prime\prime}+(2\lambda+1)(6\lambda+1)(4{A^{\prime\prime}}^2+A^{\prime\prime\prime\prime}).
\end{eqnarray}
In the last equation, the higher derivative terms drop for $\lambda=-\frac{1}{2}$ or $\lambda=-\frac{1}{6}$. At the third order in $\sfrac{1}{m^2}$, while the choice $\lambda=-\frac{1}{2}$ gives
\begin{equation}
		 \sigma{A^{\prime}}^4A^{\prime\prime}\geq0,
\end{equation}
the other choice $\lambda=-\frac{1}{6}$ leads to an inequality with higher derivative terms which we do not depict here. Therefore, from the discussion so far one can conclude that the full theory might admit a holographic c-function for $\lambda=-\frac{1}{2}$. For this choice, the NEC for the full theory defined in (\ref{3DBIgeneral}) gives
\begin{equation}
	-\frac{\sigma A^{\prime\prime}}{\sqrt{m^2+\sigma{A^{\prime}}^2}}\geq0,\label{NECBI}
\end{equation}
which has remarkably no higher derivative terms, and we can define the c-function\footnote{As shown in \cite{Gullu:2010st}, it is possible to find other functions which can be shown to be monotonically increasing when subjected to the condition (\ref{NECBI}). However, our choice here leads to a central charge which gives the central charge of NMG in the $(\sfrac{1}{m^2})$-expansion.}
\begin{equation}
	c(r)=\frac{\sigma}{\ell_P A^\prime} \sqrt{1+\frac{\sigma}{m^2}{A^{\prime}}^2},\label{eq34}
\end{equation}
which is monotonically increasing since
\begin{equation}
c^\prime(r)=-\frac{\sigma}{\ell_P {A^{\prime}}^2}\frac{1}{ \sqrt{1+\frac{\sigma}{m^2}{A^{\prime}}^2}}A^{\prime\prime}\geq0,
\end{equation}
 thanks to the condition (\ref{NECBI}).
The AdS metric with $A(r)=\frac{r}{L}$ is a solution for $\phi(r)=0$ if the vacuum equation is satisfied as
\begin{equation}
\sqrt{1+\frac{\sigma}{m^2L^2}}=1-\frac{\Lambda_0}{2m^2}.\label{BINMGvacuum}
\end{equation} 
The central charge can by read from (\ref{3DBIgeneral}) as \cite{Gullu:2010st,Nam:2010dd}
\begin{equation}
	c=\frac{\sigma L}{\ell_P} \sqrt{1+\frac{\sigma}{m^2 L^2}}\label{BIc}\,.
\end{equation}
Note that we recover equations (\ref{NMGvacuum}-\ref{NMGc}) by expanding equations (\ref{BINMGvacuum}-\ref{BIc}) in powers of $(\sfrac{1}{m^2})$.
\section{$D\geq3$ BI Gravity}\label{sec:BI}
We now study the generic BI Gravity \cite{Gullu:2014gza} theory  defined by the action given in (\ref{Generic_BI}). In order to see if it admits a c-function, we again study the Lagrangians in the $(\sfrac{1}{m^2})$-expansion by using the $D$-dimensional version of our ansatz (\ref{dwall3dwithB}), which is given by
\begin{equation}
ds^2=e^{2A(r)}\eta_{ab} dx^a dx^b+e^{2B(r)}dr^2,\label{dwallgeneral}
\end{equation}
where $\eta_{ab}$ is the Minkowski metric and the Latin indices run from $0$ to $(D-2)$. For this metric, the non-zero components of the Ricci tensor are
\begin{eqnarray}
	R^r_r&=& (D-1)e^{-2B}(-A^{\prime\prime}+A^{\prime}B^{\prime}-{A^{\prime}}^2),\nonumber \\
	R^a_b&=& e^{-2B}(-A^{\prime\prime}+A^{\prime}B^{\prime}-(D-1){A^{\prime}}^2).\label{components}
\end{eqnarray}
We can use the components (\ref{components}) to find the effective action for the functions $(A(r),B(r),\phi(r))$ in $D$-dimensions. Note that the Weyl tensor vanishes for the metric (\ref{dwallgeneral}) since it is conformally flat, and hence the terms with the Weyl tensor do not contribute to the effective action. This also means that the parameters $a_1$ and $a_2$ remain free in this context. We will not give the explicit form of the inequalities arising from the NEC since they are very complicated for the generic choice of the parameters and therefore not very illuminating. From the third and the fourth terms in the $(\sfrac{1}{m^2})$-expansion of the effective action, we find the following conditions for the cancellation of the higher derivative terms
\begin{eqnarray}
0&=& 12 a_4 (\beta +1)-4 \beta ^3-9 \beta ^2-6 \beta +\frac{D (D (7
	D-18)+12)}{(D-2)^3},\nonumber \\
0&=& 48 a_4^2 (-2+D)^5-96 \beta ^2 \left(12+20 \beta +9 \beta ^2\right)+16 D \beta  \left(-36+138 \beta +284 \beta ^2+135 \beta
^3\right)\nonumber \\
&&+ D^5 \left(11-18 \beta +3 \beta ^2+40 \beta ^3+24 \beta ^4\right)+12 D^3 \left(11-68 \beta +38 \beta ^2+164 \beta ^3+87 \beta ^4\right)\nonumber \\
&&-2 D^4 \left(37-114 \beta +27 \beta ^2+224 \beta ^3+126 \beta ^4\right)- 8 D^2 \left(9-144 \beta +192 \beta ^2+532 \beta ^3+267 \beta ^4\right) \nonumber \\
&&-24 a_4^2 (-2+D)^3 (1+\beta ) \left(D (4-16 \beta )+16 \beta
+D^2 (-1+4 \beta )\right),
\end{eqnarray}
which are satisfied by the remarkably simple choice of the parameters
\begin{equation}
a_4=\frac{D(D-1)}{2 (D-2)^2},\qquad \qquad\ \beta=\frac{D}{D-2}.
\end{equation}
With these choices, the reduced theory can be recast in a much simpler form with the help of the Schouten tensor ${\mbox{\sf P}}_{\mu\nu}$ defined as
\begin{equation}
{\mbox{\sf P}}_{\mu\nu}=\frac{1}{D-2}\left(R_{\mu\nu}-\frac{1}{2(D-1)}g_{\mu\nu}R\right).
\end{equation}
 Now, the $A_{\mu\nu}$ tensor becomes
\begin{align}
A_{\mu\nu}= & 2(D-1) {\mbox{\sf P}}_{\mu\nu}+\frac{\gamma}{m^2}\left(a_{1}\mathcal{W}_{\mu\nu}+(D-2)a_{2}C_{\mu\rho\nu\sigma}{\mbox{\sf P}}^{\rho\sigma}
+(D-1)^2{\mbox{\sf P}}_{\mu\rho}{\mbox{\sf P}}_{\nu}^{\rho}\right)\nonumber \\
& +\frac{\gamma}{Dm^2}g_{\mu\nu}\left(\frac{\left(D-1\right)^{2}}{4\left(D-2\right)\left(D-3\right)}
-a_{1}\right)\mathcal{W},\label{A}
\end{align}
which does not have any term with the trace of the Schouten tensor  $\mbox{\sf P}=g^{\mu\nu}{\mbox{\sf P}}_{\mu\nu}$.

We can now check if the full theory admits a c-function. The NEC can be derived by using the non-zero components of the Ricci tensor (\ref{components}) in the full form of the action (\ref{Generic_BI}) and the result is
\begin{equation}
	0\leq -\left(2-(D-1)\frac{\gamma}{m^2}{A^{\prime}}^2\right)^{D-2}
	\left(2(D-2)+D(D-1)\frac{\gamma}{m^2}{A^{\prime}}^2\right)A^{\prime\prime},\label{NECD}
\end{equation}
which does not have any higher derivative terms. This condition is of the following form
\begin{equation}
	\sum_{n=0}^{D-1} \left(\frac{1}{m^2}\right)^n a_n {A^{\prime}}^{2n}A^{\prime\prime}\geq0,\label{an}
\end{equation}
and the function\footnote{For $D=3$, $a(r)$ becomes the c-function $c(r)$, but for $D\geq3$ we denote it as $a(r)$  since it gives the coefficient of the A-type Weyl anomaly at the fixed point for the case of even dimensional boundary field theories.}
\begin{equation}
	a(r):=\frac{1}{({\ell_P{A^\prime})}^{(D-2)} } \sum_{n=0}^{D-1} \left(\frac{1}{m^2}\right)^n b_n {A^{\prime}}^{2n}\label{kebap49}
\end{equation}
is monotonically increasing when the coefficient $b_n$ is chosen as
\begin{equation}
	b_n=\frac{a_n}{2-D+2n}.\label{bnan}
\end{equation}
Monotonicity is clear since the derivative of the function $a(r)$ becomes
\begin{equation}
	a^\prime(r)=\frac{1}{\ell_P^{(D-2)} {A^\prime}^{(D-1)}} \sum_{n=0}^{D-1} \left(\frac{1}{m^2}\right)^n a_n {A^{\prime}}^{2n}\geq0\label{aprime},
\end{equation}
which is automatically satisfied in even dimensions thanks to the NEC (\ref{NECD}). In odd dimensions, we also need $A^\prime>0$ to ensure (\ref{aprime}). As explained \cite{Myers:2010tj}, this is realized by construction. We know that $A(r)=\frac{r}{L}$ as $r\rightarrow \infty$, which gives  $A^\prime=\frac{1}{L}>0$.  If we assume that $A^\prime<0$ for $r<r_1$ and $A^\prime>0$ for $r>r_1$ where $r_1$ is a certain value of the $r$-coordinate, then we should have $A^\prime(r_1)=0$ and $A^{\prime\prime}(r_1)>0$. However, inserting $A^\prime(r_1)=0$ into the NEC (\ref{NECD}) gives $A^{\prime\prime}(r_1)<0$, which contradicts with our assumption. Therefore, the function $A^\prime(r)$ cannot change its sign, and it satisfies $A^\prime>0$ for all values of $r$, which makes (\ref{kebap49}) well-defined.

The vacuum equation is obtained with the AdS metric for $A(r)=\frac{r}{L}$ after setting $\phi(r)=0$ 
\begin{equation}
	\frac{\Lambda_{0} \gamma}{m^2} = -1+\left(1+\frac{(D-1)\gamma}{2 m^2 L^2}\right)\left(1-\frac{(D-1)\gamma}{2m^2 L^2}\right)^{D-1},\label{BIvac}
\end{equation}
with the condition $\frac{\Lambda_{0}\gamma}{m^2}\neq-1$, which we have already assumed. Whether this equation has a unique viable solution for $\frac{1}{L^2}$ is certainly not obvious but this was proven to be the case in the Appendix-C of \cite{Gullu:2015cha}.
The value of the function $a(r)$ at the UV fixed point can be found by setting\footnote{ Note that, unlike the case of BINMG, there are $(D-1)$ terms contributing to the a-function and its value at the UV fixed point.} $A(r)=\frac{r}{L}$ as
\begin{equation}
	a=\left(\frac{L}{\ell_P}\right)^{(D-2)} \sum_{n=0}^{D-1} \frac{a_n}{2-D+2n} \left(\frac{1}{m^2 L^2}\right)^n\label{afunc}.  
\end{equation}
As suggested in \cite{Myers:2010xs,Myers:2010tj},  it should match with a universal contribution to the entanglement entropy for a particular construction, which is also proportional the coefficient associated with the A-type Weyl anomaly for odd values of $D$ (even-dimensional boundary field theories). This and further questions regarding the detailed holographic properties of the $D$-dimensional BI-type theories will be addressed elsewhere but it pays to show the 3D example explicitly as we do in the next section.

\section{$D=3$ BI Gravity as a Non-minimal Extension of NMG}\label{sec:D3BI}
 In $D=3$, all the parameters of the theory in (\ref{A}) are fixed since the terms with the Weyl tensor vanish identically and one arrives at the 3D BI gravity action
\begin{equation}
S_{\text{BI}}=\frac{2m^2}{\gamma\,\ell_P}\int d^{3}x\sqrt{-g}\,\left[\sqrt{\det\left(\delta^{\mu}_\nu+\frac{4\gamma}{m^2}{\mbox{\sf P}}^\mu_{\nu}+\frac{4\gamma^2}{m^4}{\mbox{\sf P}}^\mu_\rho {\mbox{\sf P}}^\rho_\nu \right)}-\left(1+\frac{\gamma\Lambda_{0}}{m^2}\right)\right],\label{3dmassless}
\end{equation}
which describes a massive spin-2 graviton about its unique vacuum determined by 
\begin{equation}
\frac{\Lambda_{0} \gamma}{m^2} = -1+\left(1+\frac{\gamma}{ m^2 L^2}\right)\left(1-\frac{\gamma}{m^2 L^2}\right)^{2}.\label{3dmasslessvac}
\end{equation}
The effective Planck length increases from its bare value and takes the form
\begin{equation}
	\frac{1}{\ell_P^\text{eff}}=\frac{1}{\ell_P}\left(1-\frac{\gamma^2}{L^4 m^4}\right).
\end{equation}

Up to order ($\sfrac{1}{m^6}$), the action (\ref{3dmassless}) yields
\begin{equation}
S=\frac{1}{\ell_P}\int d^{3}x\,\sqrt{-g}\left[ R-2\Lambda_{0}-\frac{4\gamma}{m^2}\left( R_{\mu\nu}^2-\frac{3}{8}R^2\right)+\frac{\gamma^2}{m^4}\left(\frac{17}{12}R^3-6R_{\mu\nu}^2 R+\frac{16}{3}R_{\mu\nu}^3\right)\right],
\end{equation}
which is the cubic order modification of NMG found in \cite{Sinha:2010ai} by requiring the existence of a c-function. Therefore, the theory defined by (\ref{3dmassless}) provides a non-minimal BI-type extension of NMG. It is non-minimal in the sense that there exists a simpler extension given by (\ref{3DBI}) defined by only the Einstein tensor and no quadratic terms in the curvature.  Black hole solutions of this cubic theory were studied in \cite{Ghodsi:2010ev}. However, the solutions for the full theory given in (\ref{3dmassless}) is an open problem.

The NEC (\ref{NECD}) in this case yields
\begin{equation}
	\left(-1-\frac{2\gamma}{m^2}{A^{\prime}}^{2}+\frac{3\gamma^2}{m^4}{A^{\prime}}^{4}\right)A^{\prime\prime}\geq0,
\end{equation}
which gives the coefficients in (\ref{an}) as $a_0=-1$, $a_1=-\frac{2\gamma}{m^2}$ and $a_2=\frac{3\gamma^2}{m^4}$.
Using these coefficients in (\ref{afunc}), one arrives at the c-function
\begin{equation}
	c(r)=\frac{1}{\ell_P A^\prime}\left(1-\frac{\gamma}{m^2}{A^\prime}^2\right)^2,
\end{equation}
and setting $A(r)=\frac{r}{L}$ one obtains the central charge
\begin{equation}
c=\frac{L}{\ell_P}\left(1-\frac{\gamma}{m^2 L^2}\right)^2.\label{3dcent}
\end{equation}
As our goal was somewhat different in this work, we have not discussed the bulk-boundary unitarity issues: but let us just note that the theory is not unitary simultaneously in the bulk and at the boundary. In fact, no massive spin-2 theory that comes from an action in 3D can be both bulk and boundary unitary as was proven in \cite{Gullu:2010vw}.

Before  we end this section, using the methods described in \cite{Henningson:1998gx,Emparan:1999pm}, we can also show that the central charge exactly matches the one obtained via the Weyl anomaly computation. For this purpose, consider the Euclidean metric
\begin{equation}
	ds^2=\frac{dr^2}{1+ \frac{r^2}{L^2}}+r^2 (d\theta^2+\sin^2\theta d\phi^2),
\end{equation}
for which the action (\ref{3dmassless}) yields\footnote{With the normalization of Brown and Henneaux \cite{Brown:1986nw}, one has $S=\frac{c}{3}  \ln\left(\frac{2R}{L}\right)$} 
\begin{equation}
	S_{\text{BI}}=\left(\frac{2L}{\ell_P}\right)\left(3+\Lambda_0L^2-\frac{3\gamma}{m^2L^2}+\frac{\gamma^2}{m^4L^4}\right)\ln\left(\frac{2R}{L}\right)\equiv 2\,c \ln\left(\frac{2R}{L}\right),
\end{equation}
where we have removed a quadratic divergence in the large cut-off scale $R$. Upon using the vacuum equation (\ref{3dmasslessvac}), one finds the same value for the central charge given in (\ref{3dcent}). Therefore, the coefficient of the Weyl anomaly can be identified with the central charge $c$ as expected.

As a further independent check, the central charge can be computed by the formula \cite{Saida:1999ec,Kraus:2005vz}
\begin{equation}
	c=\frac{L}{3\ell_P}\left(\frac{\partial\mathcal{L}}{\partial R_{\mu\nu}}g_{\mu\nu}\right)_{\bar{R}},
\end{equation}
where bar means that the quantity is to be evaluated in AdS. A straightforward computation yields the central charge given in (\ref{3dcent}).

Generic black hole solutions of the theory need to be explored, but since a Banados-Teitelboim-Zanelli black hole \cite{Banados:1992wn} can be locally identified with the AdS$_3$ spacetime, it is a solution of the 3D BI theory for the values of the couplings given in (\ref{3dmasslessvac}). The Wald entropy for the generic $D$-dimensional BI gravity was computed in \cite{Ozen:2017uoe}, and it was shown to obey the area law. For $D=3$, it takes the following particularly simple form:
\begin{equation}
	S_{W}=\frac{A_H}{4\pi}\left(1-\frac{\gamma}{m^2L^2}\right)^2.
\end{equation}
In accordance with Cardy's formula, the entropy is proportional to the central charge (\ref{3dcent}) as.
\begin{equation}
		S_{W}=\frac{A_H}{4\pi} \frac{c\,\ell_P}{L}.
\end{equation}
\section{Conclusions}
General relativity  has two crucial  properties: the existence of a unique maximally symmetric vacuum and the existence of a massless unitary graviton as the only excitation about this vacuum. Following a bottom-up approach, we can impose these properties as conditions for any viable low energy quantum gravity, but most modified theories of gravity have different particle content and vacuum structure from  general relativity, with or without a cosmological constant. However,  BI gravity, very non-trivially satisfies these  {\it a priori} very difficult conditions  with four free parameters for $D>3$. Here, in this work, we have shown that two of those parameters are fixed by the requirement of the existence of a holographic c-functions. In fact, $D=3$ theory is much simpler, and one ends up with no free parameters due to the vanishing of the Weyl tensor. We have studied this case and shown the matching of the Weyl anomaly coefficient with the central charge obtained from the holographic RG flow. In $D=4$, which we shall study in more detail elsewhere, one is left with only one free parameter $a_2$ and the defining $A$-tensor takes the form
\begin{equation}
A_{\mu\nu}=  6{\mbox{\sf P}}_{\mu\nu}+\frac{\gamma}{m^2}\left(2a_{2}C_{\mu\rho\nu\sigma}{\mbox{\sf P}}^{\rho\sigma}
+9{\mbox{\sf P}}_{\mu\rho}{\mbox{\sf P}}_{\nu}^{\rho}+\frac{9}{32} g_{\mu\nu} C_{\rho\sigma\alpha\beta}C^{\rho\sigma\alpha\beta}\right).
\end{equation}
  The rather unexpected bonus is that the BI-type theories behave as a generating function of infinitely many higher derivative theories that admit a c-function.  From these theories, one can construct a gravity theory with a c-function that has only a single massless graviton in its particle spectrum, which constitutes a useful model for testing various ideas in holography.

In the final form of the theory, given in (\ref{A}), the appearance of the Schouten tensor is not unexpected. This can be heuristically seen as follows: the Riemann tensor decomposes as
\begin{equation}
	\text{Riem}=C+\mbox{\sf P}\, \KN g,
\end{equation}
where $C$ is the Weyl tensor and $\mbox{\sf P}\, \KN g$ is the Kulkarni-Nomizu product of the Schouten tensor with the metric tensor. For deformations of a conformal metric (such as an AdS space which appear in the UV and in the IR limits), the Weyl tensor vanishes identically and the relevant non-trivial contribution comes from the Schouten part. 

Another physical constraint for the  Born-Infeld theories presented here
is the ``causality" constraint whose detailed study was initiated in \cite{Camanho:2014apa}
for the Einstein-Gauss-Bonnet theories. In the case of 3D BI gravity, it
was found in \cite{Edelstein:2016nml} that causality and unitarity are compatible
and no new condition arises from the unitarity constraint. For general BI
gravities, this is an outstanding problem on which we shall report on a
separate work.

Another interesting point to understand is the relation between the quasi-topological gravity studied in \cite{Myers:2010xs,Myers:2010tj} and the third order theory obtained from BI gravity studied here. In \cite{Gullu:2010vw}, vacuum and particle content of the quasi-topological gravity was studied and it was found that the theory is unitary in certain parameter regions.  Analogous to the relation between NMG and BINMG, one can expect a relation between the general BI theory for $D>3$ and the quasi-topological gravity.

%more on NMG, counter term, quasitopological gravity, 4d born infeld, a function c function issues, higher dimensions vs 3d, BI vacuum denklemi

\begin{acknowledgments}
G.A. thanks G{\"o}k\c{c}en Deniz {\"O}zen for helpful discussions and METU DOSAP program for support.
\end{acknowledgments}

\end{document}